\documentclass{article}

\usepackage[english]{babel}
\usepackage[letterpaper,top=2cm,bottom=2cm,left=3cm,right=3cm,marginparwidth=1.75cm]{geometry}
\usepackage{amsmath}
\usepackage[hyphens]{url}
\usepackage{longtable}
\usepackage{appendix}
\usepackage{subcaption}
\usepackage{float}
\usepackage{graphicx}
\usepackage{array}
\usepackage{authblk}
\usepackage[version=4]{mhchem}
\usepackage{siunitx}
\usepackage[utf8]{inputenc}
\usepackage[colorlinks=true, allcolors=blue]{hyperref}
\DeclareUnicodeCharacter{2032}{\ensuremath{'}}

\title{The Illusion of Understanding: How Middle-Schoolers Fail to Regulate Inquiry with ChatGPT in a Science Task}

\author[1]{Rania Abdelghani~\thanks{Corresponding author: rania.abdelghani@uni-tuebingen.de}} 
\author[1]{Kou Murayama}
\author[2]{Celeste Kidd}
\author[3]{Hélène Sauzéon}
\author[3]{Pierre-Yves Oudeyer}

\affil[1]{Hector Research Institute, University of Tübingen, Germany}
\affil[2]{Department of Psychology, University of California Berkeley, USA}
\affil[3]{Inria Research center, University of Bordeaux, France}

\begin{document}
\maketitle

\begin{abstract}
Generative AI (GenAI) tools allow for effortless task completion, potentially fostering cognitive and metacognitive laziness in students. While surveys indicate widespread GenAI use among students as young as 11, their interactions strategies remain under-explored.

A critical indicator of these interactions' quality is the ability to lead Question-Asking (QA) cycles: initiating goal-oriented inquiries, critically evaluating AI responses, and regulating subsequent strategies. While these behaviors predict robust learning in traditional settings, their role in AI-mediated environments remains unclear.

Addressing this gap, this study investigates middle school students' (N=63, aged 14--15) capacity to adopt these behaviors with GenAI during science investigation tasks. We analyzed their proficiency in distinguishing efficient goal-oriented prompt from inefficient ones, their critical evaluation of AI responses, and their ability to generate follow-up questions to regulate learning in alignment with their informational needs.

Findings reveal a pattern of over-reliance: students struggled to discriminate between prompt types, failed to detect vague AI explanations, and frequently terminated inquiry prematurely, without follow-up. Consequently, task performance remained moderate despite unrestricted AI access and high self-reported prior knowledge. Notably, positive AI attitudes were negatively associated with interaction quality, suggesting a disconnect between perceived and actual competence, whereas higher metacognitive skills predicted superior sensitivity to prompt quality.

These results underscore the necessity for AI literacy interventions that move beyond technical understanding to explicitly train metacognitive regulation strategies, required for meaningful and sustainable QA-based learning with GenAI.
\end{abstract}


\section{Introduction and Related Work}
\label{sec:intro}
\paragraph{Learning Challenges in the GenAI Era and the Role of Question-Asking (QA)}
Generative AI (GenAI) tools, particularly Large Language Models (LLMs), are increasingly being used in education, with recent reports indicating that most students already rely on them for school-related tasks~\cite{commonsense,microsoft,hashem2025understanding}. While they present exciting opportunities to tackle challenges in educational research~\cite{kasneci2023}, they also come with their own unique challenges. Indeed, due to their lack of pedagogical alignment~\cite{sonkar2024pedagogical}, they can come with several risks that may hinder learning: promote cognitive and metacognitive laziness~\cite{fan2025beware}, over-reliance issues~\cite{zhai2024effects}, and even the adoption of mis- or biased information~\cite{kidd2023}.

During interactions with these tools, students use natural language to formulate what is called a “prompt”, a process that fundamentally draws on their question-asking (QA) skills~\cite{sasson2025art}. Although the field is still emerging, preliminary research suggests that the pedagogical effectiveness of student–LLM interactions may heavily depend on students’ ability to formulate clear, context-specific, and well-structured questions, as well as to critically evaluate the model’s responses~\cite{kasneci2023,kong2024developing,chiu2024define}. Beyond reducing the risk of AI misbehavior~\cite{aydin2023chatgpt}, these high-quality QA-based strategies are believed to promote greater cognitive engagement, thereby helping to mitigate passivity and over-reliance on LLMs during learning~\cite{abdelghani2023generative,ruggeri2016sources}. 

It is thus suggested that the crux of the challenge of promoting efficient use of AI during learning can rely on the students' ability to engage in efficient QA behaviors with the tool. 

\paragraph{Characterizing Meaningful QA during Prompting}
To characterize these behaviors, we rely on QA taxonomies resulting from research in cognitive science, focusing on QA skills in standard learning environments. This research suggests that, in order to be efficient for learning, QA behaviors must be "meaningful"—engaging optimal cognitive resources to refine one's understanding and world model, rather than merely extracting isolated facts~\cite{graesser1993anomalous}. From an information-seeking perspective, meaningful QA is not a simple state but a dynamic cycle requiring sophisticated cognitive skills, including metacognitive control, flexibility, and conceptual knowledge of the QA process~\cite{ruggeri2016sources,abdelghani2023interactive}. Specifically, this behavior entails: 1) \textit{Monitoring} one's knowledge state to specific epistemic gaps (informational needs) that, when filled, will advance understanding~\cite{chouinard2007children}; 2) \textit{Initiating} the search to bridge these gaps by formulating clear, contextually precise queries that accurately convey informational needs~\cite{ronfard2018question}; and 3) \textit{Evaluating} incoming answers against initial goals to regulate subsequent learning strategies (e.g., verifying, clarifying, or integrating)~\cite{murayama2019process,Oudeyer2018}.

Mastery of these behaviors is a robust predictor of academic success and is significantly associated with positive affective experiences, such as curiosity and learner agency~\cite{jirout2011curiosity}. Research in developmental science shows that children, by the age of 7, can select meaningful questions to compensate for their informational needs, and can formulate their own by 10~\cite{mills2011determining}. At this age, they can also recognize unsatisfying answers and require further information~\cite{mills2019want}.

\paragraph{Challenges of Meaningful QA with LLMs} In the context of GenAI and LLMs, these high-level QA manifest through "prompting" strategies. While often perceived as effortless, effective prompting mirrors the foundational components of meaningful QA~\cite{sasson2025art}: it requires users to articulate clear goals, formulate clear context, and critically verify outputs~\cite{denny2024prompt}. Consequently, "prompt engineering" in education can be conceptualized as a manifestation of meaningful QA skills in LLM-based contexts~\cite{kasneci2023}. 

However, implementing these behaviors with LLMs presents unique challenges. Unlike human tutors, who typically scaffold the QA process by requesting clarification for vague inquiries, providing feedback on formulation, or validating understanding after giving answers~\cite{cheng2025asking}, LLMs lack such inherent pedagogical grounding~\cite{sonkar2024pedagogical}. They often provide direct, confident answers to literal prompts—even when ambiguous or have low quality and pedagogical value—failing to guide the learner towards more precise formulations~\cite{aydin2023chatgpt,deng2024don}. This lack of scaffolding shifts the entire regulatory burden of the QA process onto students, a task particularly difficult for novices who are still developing these skills~\cite{tankelevitch2024metacognitive,ruggeri2016sources}. In this scenario, students must manage not only the traditional cognitive and metacognitive demands of QA but also navigate the additional unique complexities inherent to LLM interactions~\cite{tankelevitch2024metacognitive,zhai2024effects}. 

Furthermore, the perceived "friendliness" and ease of LLMs appear to discourage germane cognitive effort. Research suggests that learners tend to delegate essential metacognitive functions to the AI~\cite{fan2025beware,zhai2024effects}, bypassing the self-reflection needed to identify knowledge gaps in favor of vague queries. This aligns with findings that children invest less effort in question formulation when they perceive their informant as "friendly" and "not judgmental"~\cite{ronfard2018question}. Empirical evidence reinforces this concern: students using LLMs report lower germane cognitive load (essential for deep learning) and achieve poorer learning outcomes compared to those using standard search engines~\cite{stadler2024cognitive}. Similarly, it is suggested in~\cite{Yan2025DistinguishingAI,stadler2024cognitive}, that students would tend to delegate metacognitive regulation to the AI, failing to accurately identify their informational needs or seek and evaluate information efficiently.

Such deficiencies in formulation and evaluation likely result in passive learning, superficial understanding, and increased susceptibility to misinformation~\cite{kidd2023}. Strategies that foster meaningful QA are therefore essential to mitigate over-reliance and sustain learner agency with GenAI~\cite{saadat2024not}. Investigating the current reality of these behaviors is a foundational first step towards implementing such strategies. This understanding is crucial for informing the design of evidence-based interventions or integrating QA-scaffolding strategies into educational LLMs to promote active and critical engagement~\cite{darvishi2024impact}.

\paragraph{Empirical Evidence: Students Struggle with LLM Interactions} In this context, existing empirical evidence suggests indeed that students currently struggle to maintain meaningful QA cycles with LLMs. Regarding \textit{prompt formulation}, studies indicate that even university students frequently favor quick, copy-paste prompts over deep exploratory questions~\cite{cheng2025asking}. For instance, computer science graduates have been shown to struggle with articulating effective questions for code generation, leading to low success rates~\cite{denny2024generative,babe2023studenteval}. Similarly, non-expert adults often fail to test different questions or overgeneralize prompt characteristics~\cite{zamfirescu2023johnny}. Regarding \textit{answer evaluation}, students often accept AI responses without critical assessment. Hill et al.~\cite{hill2023} demonstrated that students with access to ChatGPT performed worse on exams than those without, largely due to uncritical reliance on inaccurate answers. This passivity is exacerbated by the difficulty of detecting LLM-generated misinformation, which non-experts find harder to identify than human-generated errors~\cite{chen2023can}. Crucially, research in this area predominantly focuses on higher education, leaving a significant gap regarding younger learners (K-12) despite highly-rising adoption rates~\cite{du2024factors,hashem2025understanding,sheese2024patterns}.

To address this gap, this study investigates French middle school students' (aged 14--15) capacity to initiate and evaluate meaningful QA cycles with ChatGPT during science investigation tasks. We focus on science as it represents a core curriculum subject that inherently relies on inquiry skills—generating hypotheses, testing them through targeted questioning, and critically evaluating evidence~\cite{zoller1997student}—yet remains underexplored in LLM research compared to coding~\cite{lo2023impact}. We argue that investigating this age range is critical, as early adolescence is a sensitive developmental window for establishing both metacognitive strategies and digital habits~\cite{ruggeri2016sources,fandakova2021states}. The challenges of meaningful QA are arguably more pronounced for this cohort: younger students generally display a higher tendency to trust AI-generated content, placing them at significant risk of implementing sub-optimal QA behaviors that impede learning quality~\cite{kidd2023,noles2015children,van2021toy}. 

\section{Current study}
Grounded in cognitive science research, we investigate two main components of meaningful QA~\cite{ronfard2018question}:
\begin{itemize}
    \item \textbf{Initiation of the QA cycle (Prompt Discrimination)} This refers to students' ability to discern "efficient" prompts for initiating task exploration. Efficient prompts are clear, goal-oriented and contextually-sufficient. They are more likely to lead to meaningful QA cycles by generating information that requires deep cognitive processing and advances the learning of the task. To measure this, students are presented with suggestions for efficient versus inefficient prompts for each task and asked to judge their quality. They are then free to use this suggestion or formulate their own prompt.
    
    As discussed above in~\autoref{sec:intro}, this design is theoretically grounded in developmental research suggesting that the competence to \textit{identify} effective questions typically precedes the ability to independently \textit{formulate} them~\cite{mills2011determining}. This distinction also allows us to methodologically isolate the judgment component of prompting skills, providing a controlled metric for our QA-initiation skills analysis See~\autoref{sec:study-design} for more details about the design.
    
    \item \textbf{Evaluation and regulation of the QA cycle (AI Answer Evaluation \& Follow-up} This refers to students' ability to critically assess the epistemic utility of AI responses and strategically regulate their QA process (e.g., by requesting further questions or terminating the search). 
    
    To assess this, every AI-generated response encountered during exploration is evaluated by domain experts for cognitive depth and meaningfulness, following Graesser's taxonomy~\cite{graesser1994question}. We then compare students' subjective utility ratings against this expert to determine their sensitivity to answer quality. We also  assess regulatory strategies in LLM environments by analyzing the tendency to initiate follow-up questions versus terminating the QA cycle prematurely. This serves as a indicator for persistence and regulatory skills in LLM environments.
    \end{itemize}

Beyond characterizing these behaviors, we draw upon research into the determinants of meaningful QA in traditional settings to examine whether they play a similar role in LLM-mediated environments. Specifically, we analyze the influence of metacognitive abilities~\cite{metcalfe2020epistemic}, perceptions of the informant (the AI in this case)~\cite{ronfard2018question}, and prior domain knowledge~\cite{ruggeri2016sources}. Finally, we explore the predictive relationship between these QA-based interaction behaviors and task outcomes, defined as the students' ability to explain scientific concepts in their own words.

With this work, we address the following questions:
\begin{enumerate}
\item \textbf{RQ1 (Prompting -- QA initiation):} Can students efficiently discriminate between efficient QA-based prompts and inefficient ones?
\item \textbf{RQ2 (AI Answer Evaluation -- QA evaluation):} Can students efficiently assess the quality and utility of GenAI’s outputs, with respect to their learning goals?
\item \textbf{RQ3 (AI Answer Follow-Up -- QA regulation):} Can students strategically regulate their QA strategies (e.g., follow-up vs. termination of QA cycle) upon receiving AI responses?
\item \textbf{RQ4 (Task Outcomes)} Can LLM-supported QA-based cycles lead to successful conceptual understanding of scientific concepts? 
\item \textbf{RQ5 (Determinants of QA with GenAI)} How do individual differences—prior domain knowledge, AI attitudes, and metacognition relate to QA-based interaction quality and subsequent task outcomes with LLMs?
\end{enumerate}

\section{Study design}
\label{sec:study-design}
\subsection{Participants}
We recruited 73 middle school students from four institutions (two public, two private) in the Nouvelle-Aquitaine region, France. Following data cleaning—excluding 10 participants for incomplete tasks or questionnaires—the final sample comprised 63 students (32 males, 29 females, 2 non-disclosed; aged 14--15 (mean=14.07). Written informed consent was obtained from all participants and their legal guardians.

An a priori power analysis ($\alpha=.05, \quad 1 - \beta=.9$) indicated a required minimum sample of N=63 to detect a Smallest Effect Size of Interest (SESOI) of r=0.3 for relationships between interaction scores and individual differences.

\subsection{Task description}
\begin{figure}
\centering
\includegraphics[width=1\linewidth]{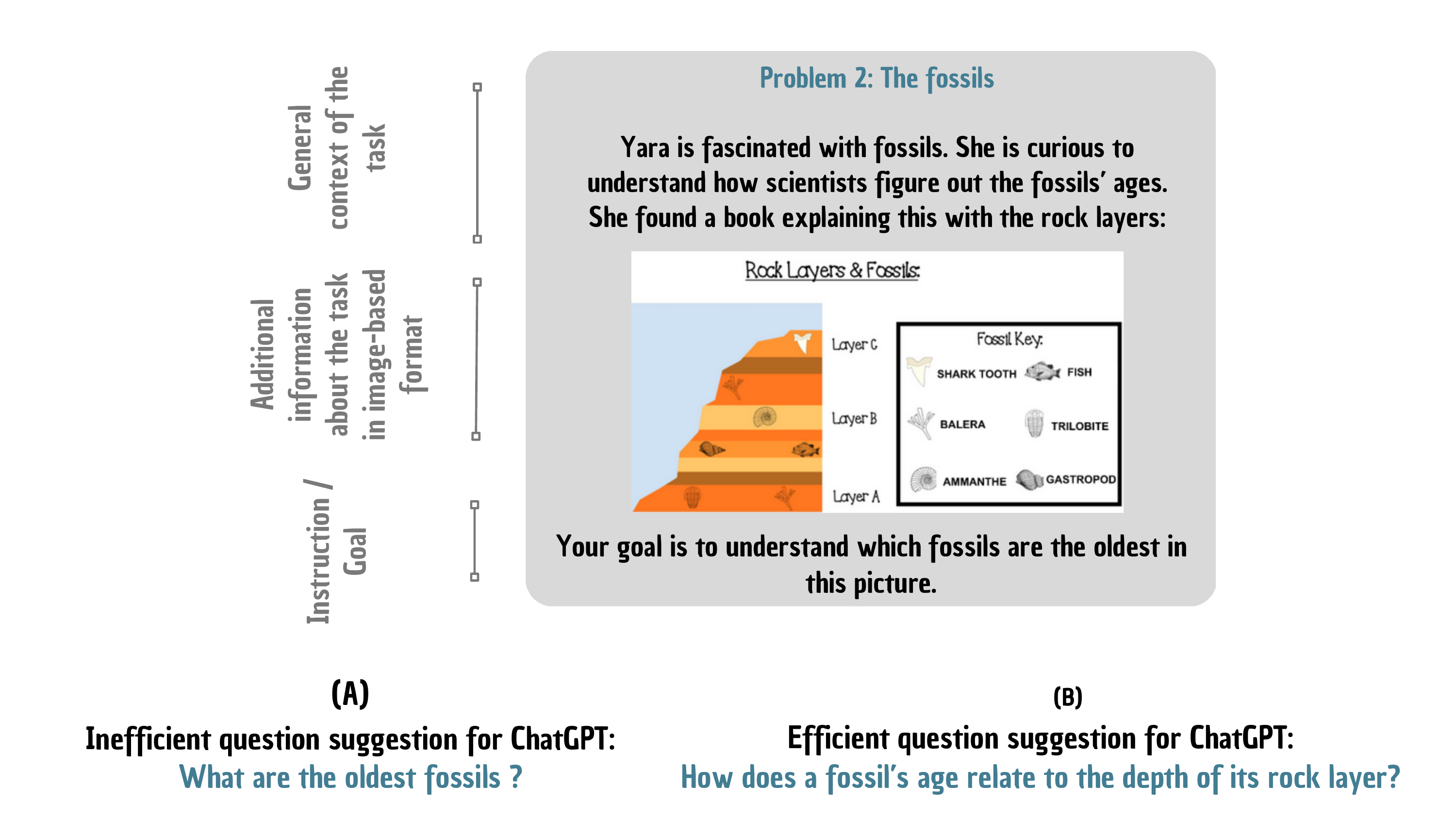}
\caption{  \centering An example of the tasks proposed. Each task includes specific informational elements presented in both text and image format to understand its specific context, along with a text-based instruction. (A) is an example for a case with a 'non-efficient' question for the prompt and (B) is for an 'efficient' one.}
\label{fig:task_sample}
\end{figure}

The study employed six science investigation problems validated by middle school science teachers for curricular relevance. Each problem comprised a textual context, a descriptive image, and a specific learning goal (see an example in~\ref{fig:task_sample}).

Crucially, each problem included a randomized suggested prompt to initiate the QA cycle with the LLM. Drawing on question taxonomies~\cite{graesser1993anomalous,chouinard2007children,alaimi2020pedagogical}, and as already discussed above, these suggestions were manipulated into two conditions to mirror the criteria for meaningful questions: 
\begin{enumerate} 
\item \textbf{Efficient Prompts: likely leading to meaningful QA cycles} Formulated with precise learning goals and context to elicit deep, explanatory responses that require high-level cognitive processing and facilitate engagement and conceptual understanding of the task.
\item \textbf{Inefficient Prompts: less likely to lead to meaningful QA cycles} Vague, superficial or context-poor formulations  likely to yield generic, isolated or incomplete AI outputs that do not advance understanding and do not require deep cognitive processing.
\end{enumerate} 

Before the study, we run several validation steps confirming that efficient prompts consistently yielded superior pedagogical and engaging content, though we control for ChatGPT's non-determinism in our analysis (see~\autoref{sec:results}). An example of the interface with an efficient vs. inefficient suggestion is shown in~\autoref{fig:task_sample}. The full pool of tasks is available in the supplement materials~\footnote{\label{smfootnote}\url{https://osf.io/ehqg9/overview?view_only=07da924434f143a9bb7902f7605229b8}}.

For each task, students first evaluated the quality of the suggested prompt. Subsequently, they could either use the suggestion or formulate their own questions to start exploring the problem with the LLM. We allowed unrestricted QA cycles, with students required to evaluate every AI response received (see~\autoref{sec:proc-measures}).

\subsection{Procedure and Measures}
\label{sec:proc-measures}
Data collection occurred in schools, in normal class hours, and in groups of 10. Researchers first briefed participants on the study goals and demonstrated an example task to ensure comprehension of the interface and protocol. Following the briefing, students completed two individual questionnaires before starting the problem-solving session.

\begin{itemize}
    \item \textbf{Experience, attitude and perceptions of GenAI} We adapted the Bernabei et al.~\cite{bernabei2023students} scale for adolescents, resulting in a 24-item instrument, rated on a a 4-point Likert scale (from 0 to 3). The questionnaire assesses six dimensions: \textit{Attitude, Trust, Social Influence, Fairness, Usefulness,} and \textit{Effort}. The overall maximum score is 72. See the supplement materials for detailed items\hyperref[smfootnote]{\footnotemark[\getrefnumber{smfootnote}]}.
        
    To check the internal consistency of this new version of the questionnaire, we calculated Cronbach's Alpha and had a good reliability result ($\alpha$ =0.8, 95\% CI=[0.73;0.86]).

    \item \textbf{Metacognitive competencies} To measure students' ability to monitor and regulate their learning, we used the Junior Metacognitive Inventory (Jr. MAI)~\cite{kim2017establishing}. Our goal being to investigate this prerequisite in LLM contexts. This instrument has 18 items (rated on a 5-point Likert scale: 0-4) and contains two sub-scales: \textit{Knowledge of Cognition} and \textit{Regulation of Cognition}. See the supplement materials for detailed items \hyperref[smfootnote]{\footnotemark[\getrefnumber{smfootnote}]}.
\end{itemize}

After completing the questionnaires, and during a one-hour session, students completed six science investigation tasks. To prevent direct copy-pasting and enforce active processing, tasks descriptions and prompt suggestions were provided on paper, while interactions with ChatGPT 3.5 (pre-loaded) and data logging occurred on laptops provided by the research team. Tasks were randomly assigned from a pool of 12 equivalent and validated exercises to prevent collaboration. 

For each task, the workflow was structured as follows: 
    \begin{enumerate} 
    \item \textbf{Self-report of Prior Knowledge:} Students read the task and the suggested prompt (efficient or inefficient). On a web platform loaded also on their laptops, they report their confidence in their prior Knowledge in the domain knowledge, using a 3-point scale: 1) not at all confident about having knowledge about this task, 2) a bit confident about having knowledge about this task, or 3) very confident about having knowledge about this task.
    
    \item \textbf{Evaluation of the Prompt Suggestion:} students use the same platform to indicate whether they intend to use the suggested prompt or not, by choosing one of the suggestions: 1) This is a good prompt that can help me explore the task efficiently with ChatGPT 2) This is not a good question to explore the task efficiently with ChatGPT.

    \item \textbf{Exploration \& Monitoring with the LLM:} Students start interacting with ChatGPT with no restrictions on the number of turns. Crucially, they were required to evaluate \textit{every} AI response received on a 3-point scale: 1) it does not at all provide useful information to help me understand the task, 2) it provides a general/incomplete information related to the task but not fully help me understand the task or 3) it provides the useful information I need for better understanding of the task.
    
    \item \textbf{Answer Submission:} The QA cycle for one task ends when students feel satisfied with the information they gained. They use the web platform to submit an answer in their own words, using a maximum of 3 sentences. 
    \end{enumerate} 
    
\begin{figure}
\centering
\includegraphics[width=1\linewidth]{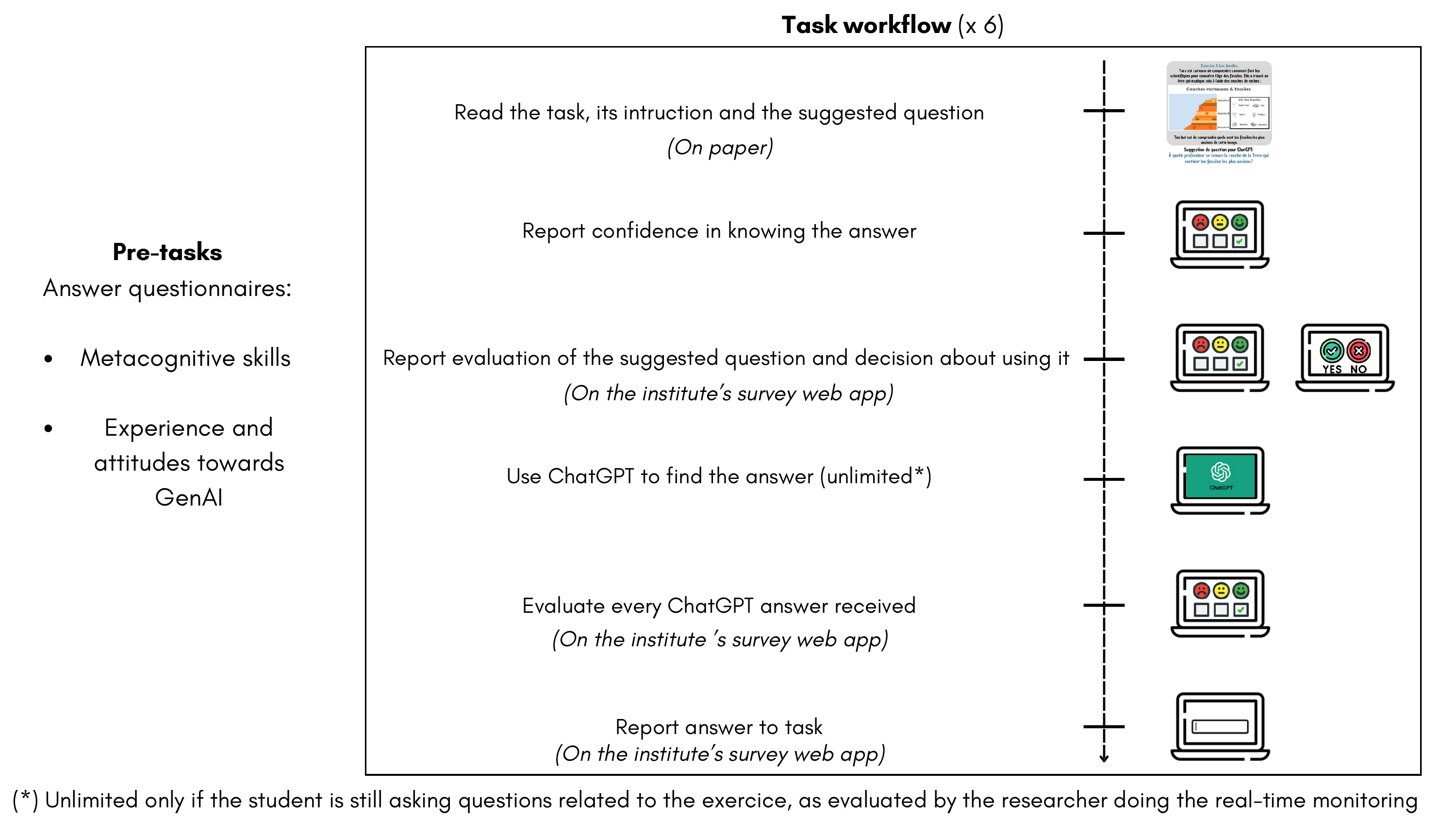}
\caption{\label{fig:study_timeline}Overall study timeline}
\end{figure}

See~\autoref{fig:study_timeline} for the complete study timeline and data collection process.

\subsection{Ethical considerations}
To ensure safety, we implemented a real-time monitoring protocol allowing experimenters to intercept queries containing Personally Identifiable Information, offensive content, or persistent off-task inputs (defined as more than 3 consecutive irrelevant queries). While experimenters reserved the right to block such inputs and intervene, no violations occurred during the study. Participants were explicitly briefed on these usage constraints prior to the session.

The study was approved by the ethical committee of the research center (COERLE). All security and risk management measures were also discussed and approved by the schools' boards where we intervened.

\section{Results}
\label{sec:results}
To address our research questions, we organize our data analysis like the following: first, we start by investigating students' ability to distinguish "efficient" and "non-efficient" prompts. Second, we run a similar analysis for their ability to distinguish useful AI answers and to ask follow-up questions. Finally, we investigate multi-level links between these interaction measures, individual differences (i.e. metacognitive skills, perceptions of AI and prior knowledge), and task outcomes.

\subsection{Initiating the QA cycle}
\paragraph{Usage Patterns}
Analysis of the interaction logs reveal a widespread reliance on our suggested prompts. In total, 94.1\% of the students used the suggested prompts, at least once for initiating their QA cycle. Examining the distribution of strategies, the majority of the cohort (45 students) adopted a hybrid approach, using both suggested prompts and formulating their own. Meanwhile, 16 students relied exclusively on the suggested options, and only 2 students relied exclusively on self-formulation. Consequently, 74.6\% of the participants attempted to initiate the exploration with their own prompts at least once across the 6 tasks.

It is to be noted here that these results refer to students' objective behavioral patterns, rather than their \textit{perception} of the quality of suggested prompts. When students explicitly judged a prompt as "efficient," they proceeded to use it 84.9\% of the time. However, a notable disconnect emerged regarding inefficient prompts: students proceeded to use prompts they had explicitly judged as "inefficient" in 75.9\% of instances. In other words, students chose to inhibit the use of perceived inefficient prompts in only 24.1\% of cases. This can suggest that while students engaged with the selection task, their usage behavior was not strictly motivated by the goal to maximize task outcomes.

\paragraph{Prompt Discrimination Skills}
To assess proficiency in judging prompt quality, we computed the Signal Detection Theory sensitivity index \textit{$d'$}~\cite{fleming2014measure}. This measure represents the Z value of the difference between the hit rate (in our case, decision to accept a suggested prompt when it is efficient) and the false alarm rate (decision to accept a suggested prompt when it is not efficient). We applied a log-linear correction ($\epsilon = \frac{0.5}{N}$); N being the total number of trails, i.e. 6) for extreme rates. Crucially, we calculated d′ based on students' \textit{evaluation} of the suggestions rather than their usage, isolating judgment from personal preference. For example, a student might choose not to use a suggested question despite recognizing its quality, simply because they prefer to formulate their own; such preferences did not affect our measure.

\begin{figure}%
    \centering
    \subfloat[\centering \textbf{Students' sensitivity to the suggested prompts' quality was low at average.} ]{{\includegraphics[width=7cm]{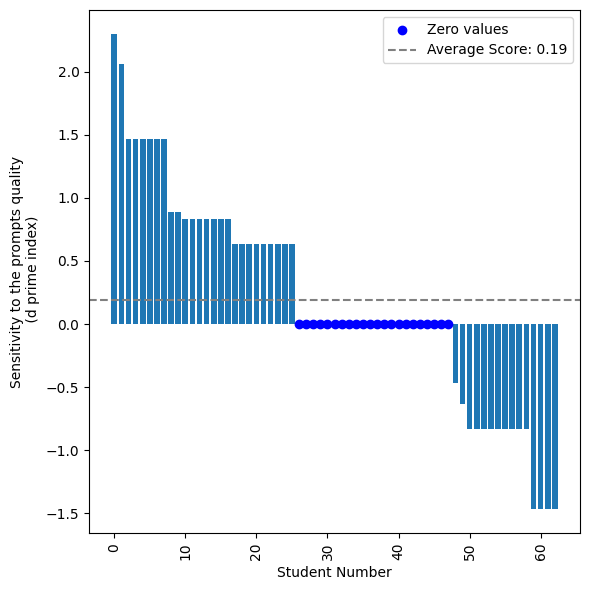} }}%
    \subfloat[\centering \textbf{Students' self-generated prompts led mainly to low-level answers}]{{ \includegraphics[width=7cm]{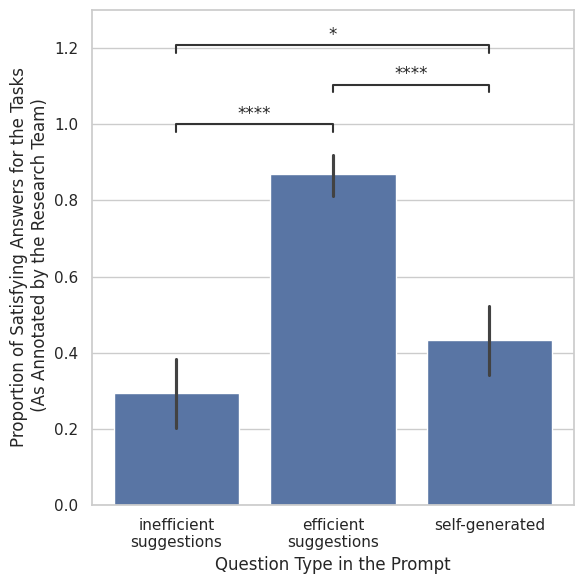} }}%
    \caption{\centering Performance during QA initiation}%
    \label{fig:res-prompt}%
\end{figure}

As illustrated in Sub-figure (a) in~\autoref{fig:res-prompt}, this measure had rather low values: mean $M_{\text{\scriptsize d$'$}}$=0.19, deviation $SD_{\text{\scriptsize d$'$}}$= 0.8. A one-sample t-test against zero confirmed that students performed at chance level (t=1.77, p=0.08), indicating a general inability to reliably distinguish between efficient and inefficient prompts. However, the measure also shows substantial individual differences, suggesting the need for further investigations.

\paragraph{Self-generated Prompts Quality}
In a second step, We subsequently assessed students' ability to \textit{formulate} meaningful questions by analyzing the quality of their corresponding resulting AI answers. Following Graesser's taxonomy~\cite{graesser1994question}, we manually binary-coded outputs as either \textit{High-Level}: information that is clear and requires deep cognitive processing such as understanding mechanisms, making inferences, etc. It helps engage in the task and better understand it. Or\textit{Low-Level}: information that only represents an isolated, vague or general piece of information, circular explanation, etc. It does not meaningfully help learn the task.

We chose to analyze answer quality as a proxy to assess the quality of questions, given the lack of coding schemes allowing the annotation of questions that can be applied to our experimental context and goals. The pedagogical value of a question is suggested to be intrinsically linked to the cognitive depth of the resulting information: meaningful questions are associated with high-level answers, whereas less efficient ones are associated with low-level information~\cite{graesser1994question,alaimi2020pedagogical,mills2019want}.

Our analysis shows that our efficient suggestions led to high-level answers in 84.4\% of the time, compared to 38.9\% for self-generated prompts and 31.1\% for inefficient suggestions. A logistic regression confirmed that prompt type significantly predicted answer validity ($<$0.0001). In doing pairwise comparisons, we see that: 1) using efficient suggestions, compared to inefficient ones significantly increased the odds of getting a high-level answer as judged by the research team (z=8.3, p$<$0.0001, 95\% CI=[2.13; 3.44]). And 2) while the self-generated prompts resulted in more valid answers than the inefficient suggestions (z=2.08, p=0.037, 95\% CI=[0.036; 1.18]), they also led to significantly less valid answers than our efficient suggestions (z=-3.9, p=$<$0.0001, 95\% CI=[-1.31; -0.44]). See sub-figure (b) in~\autoref{fig:res-prompt} for a visualization of the results.


\subsection{Evaluating the QA cycle}
\paragraph{Usage Patterns}
Students were asked to rate every answer they received from ChatGPT during their interactions, on a scale from 1 to 3, as described in~\autoref{sec:proc-measures}. We then compared these subjective evaluations with the objective quality of the answers as annotated by the research team (the same annotations mentioned in the sub-section above). In total, we had a pool of 400 questions sent to ChatGPT during this session.

Our descriptive statistics showed that students rated 71.4\% of the answers rated "\textit{low-level}" by experts as clear and useful (Scores 2 or 3), a rate nearly identical to their acceptance of high-level answers (75.9\%). Only 16 students achieved an evaluation accuracy exceeding 70\%.

\paragraph{Answer Quality Discrimination}
To quantify this lack of calibration, and similar to prompt sensitivity, we compute a sensitivity index $d'$ for answer evaluation. In this case, the index will refer to the ability to discriminate between satisfying and unsatisfying answers, treating the experts' annotations as "ground truth".

Our results show that on average, students demonstrated a near-zero sensitivity: $M_{\text{\scriptsize d$'$}}$=0.07, $SD_{\text{\scriptsize d$'$}}$= 1.2. As it was the case for the sensitivity to the questions' quality, t-test confirmed that performance was indistinguishable from chance: t=-0.45, p=0.65, thus suggesting a profound inability to discern informative from uninformative AI outputs. Also similar to our previous results, we see high inter-student variability. See sub-figure (a) in~\autoref{fig:res-output}. More specifically, our results show that students often failed to recognize the answers rated as low-level by the research team, frequently giving them high ratings. 

\subsection{Regulating the QA cycle}
Behavioral regulation was notably absent: in total, only 14 students asked follow-up questions (totaling 22 questions across 6 tasks and all students). 7 of these questions were asked after using inefficient prompts, 4 after using efficient ones, and 11 after using one's own prompt. Meaning that even in the instances where students correctly rated an answer as unsatisfactory, they initiated follow-up questions only in 31.8\% of the time, suggesting a lack of motivation to repair the QA cycle, while when seeing learning opportunities. 

\begin{figure}%
    \centering
    \subfloat[\centering \textbf{Students' sensitivity to ChatGPT answers quality was significantly low.} ]{{\includegraphics[width=6.5cm]{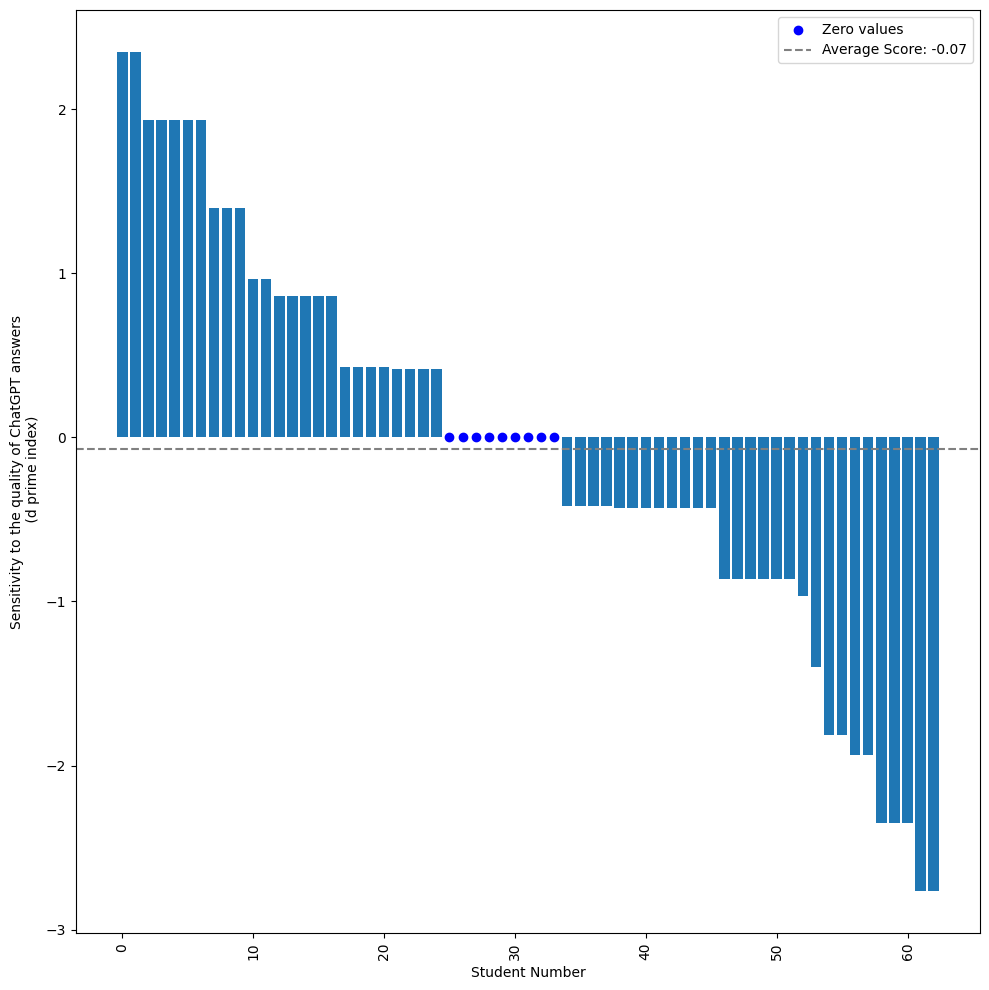} }}%
    \hfill
    \subfloat[\centering \textbf{Students assigned high ratings to low-level ChatGPT answers, even with high prior confidence in task domain.}]{{ \includegraphics[width=6.5cm]{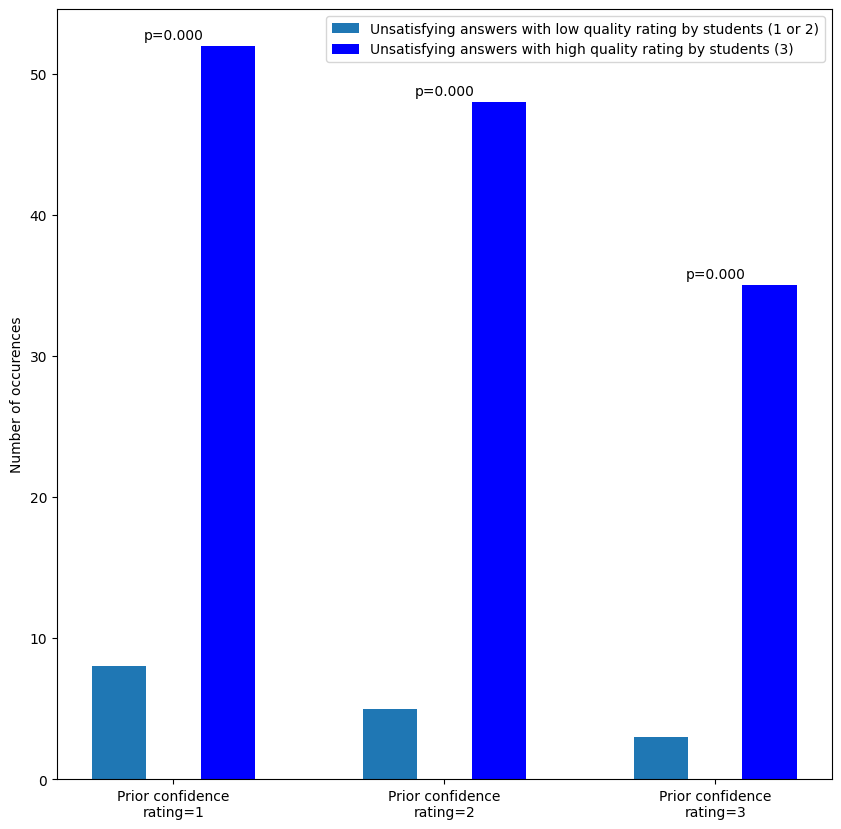} }}%
    \caption{\centering Performance during QA evaluation.}%
    \label{fig:res-output}%
\end{figure}

Crucially, this sub-optimal evaluation persisted regardless of self-reported prior knowledge confidence (see~\autoref{fig:res-output}), meaning that domain knowledge failed to shield students from the "illusion of understanding" created by GenAI fluency. This aligns with findings that younger learners are particularly impressionable and vulnerable to misinformation~\cite{kidd2023,fazio2015knowledge}.

\subsection{Resulting Task Outcomes}
Finally, we examined students’ task outcomes. They were defined as the capacity to synthesize a deep, accurate explanation of the scientific concept. We manually binary-coded final written responses: 1 if the explanation was deep, clear, explicitly addressing the problem, and accurate and 0 if not. Importantly, students were explicitly instructed to write their final solutions in their own words. To ensure scores reflected genuine conceptual understanding rather than retrieval, direct copy-pastes from ChatGPT were excluded. 

\begin{figure}
\centering
\includegraphics[width=.5\linewidth]{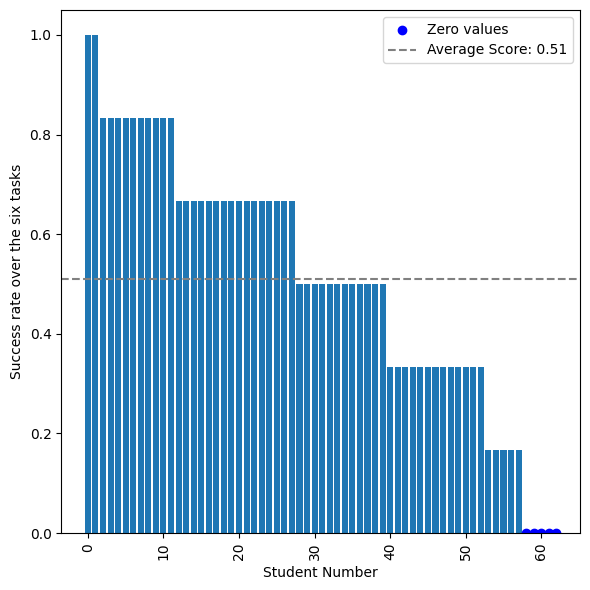}
\caption{  \centering Students had an average chance-level success rate over the six tasks.}
\label{fig:success_average}
\end{figure}

As illustrated in Figure~\autoref{fig:success_average}, the average correct solution rate over the six problems per student was of $M_success$=0.51 and $SD_follow$=0.25. This rate can be considered as relatively low, considering that all tasks were solvable with ChatGPT when prompted effectively and sustainably.

\subsection{Links with Individual Differences}
We run a path analyses to investigate how individual differences (metacognition, perceptions and knowledge of AI, and prior domain-knowledge) can influence these critical QA-based interaction behaviors with the LLM and the subsequent task outcomes.

\paragraph{Prompt Quality Sensitivity} Proficiency in discriminating efficient prompts was negatively predicted by positive AI attitudes (first sub-scale of AI questionnaire described in~\autoref{sec:proc-measures}): $\beta$=-.39, p=.015. Bi-variate correlations further indicate that students performed better when they reported lower scores on this sub-scale (r=-.49, $p<$.001), had higher critical scores of AI fairness—another sub-scale of the questionnaire (r=-.31, p=.023, reversed items), and higher metacognitive regulatory skills (r= .3, p=.022)—a sub-scale of the JrMAI questionnaire. 

This suggests that students' overconfidence in their AI knowledge and experience may hinder their judgment, whereas metacognitive regulation can support it~\cite{ruggeri2016sources}. Results of metacognitive skills are in line with standard information-search findings suggesting its role to facilitate the initiation of efficient QA cycles~\cite{ruggeri2016sources,murayama2019process}.

\paragraph{Answer Evaluation Sensitivity} Similarly, the ability to evaluate AI answers was negatively associated with AI attitudes ($\beta$=-.3, p=.018), and, notably, with prior domain knowledge, as reported by students themselves before starting the task ($\beta$=-.35, p=.042). This indicates that domain knowledge did not shield students from the "illusion of understanding"; rather, high confidence appeared to reduce epistemic vigilance.


\paragraph{Regulatory Behavior}
Active regulation was driven by critical perceptions of AI. Indeed, the frequency of generating follow-up questions was significantly related to three predictors: critical perception of AI fairness ($\beta$=-.33, p=.014), perception of AI usefulness ($\beta$=-.26, p=.05), and perception of effort needed to use AI ($\beta$=.3, p=.049)—all sub-scales of the AI questionnaire. These findings suggest that follow-up QA behaviors reflect engaged and critical interactions rather than random behavior: students who viewed the tool skeptically engaged more deeply in QA cycles than those who perceived it as an effortless, super-useful, and fair tool. Notably, we do not find direct interactions with metacognitive measures. 

These results align with previous QA research in standard settings, suggesting that students' conceptual knowledge of their informant's accuracy and malleability shapes the way they implement their inquiries~\cite{ronfard2018question,abdelghani2023interactive}. 

\paragraph{Task Outcomes}
Finally, to investigate the task outcomes as a final measure of interest, we run multiple linear regression models. The best fitting model (F(5,57)=40.29, p$<$0.0001, adjusted $R^2$=0.81) showed that the success rate across tasks was predicted by three variables: the frequency of satisfying answers received—i.e. rated as high-level by the research team: $\beta$=0.52, p=0.000; their sensitivity to AI answers quality: $\beta$=0.1, p=0.000; and the ability to ask follow-up questions after receiving low-level answers: $\beta$=0.4, p=0.004.

\begin{figure}
\centering
\includegraphics[width=1\linewidth]{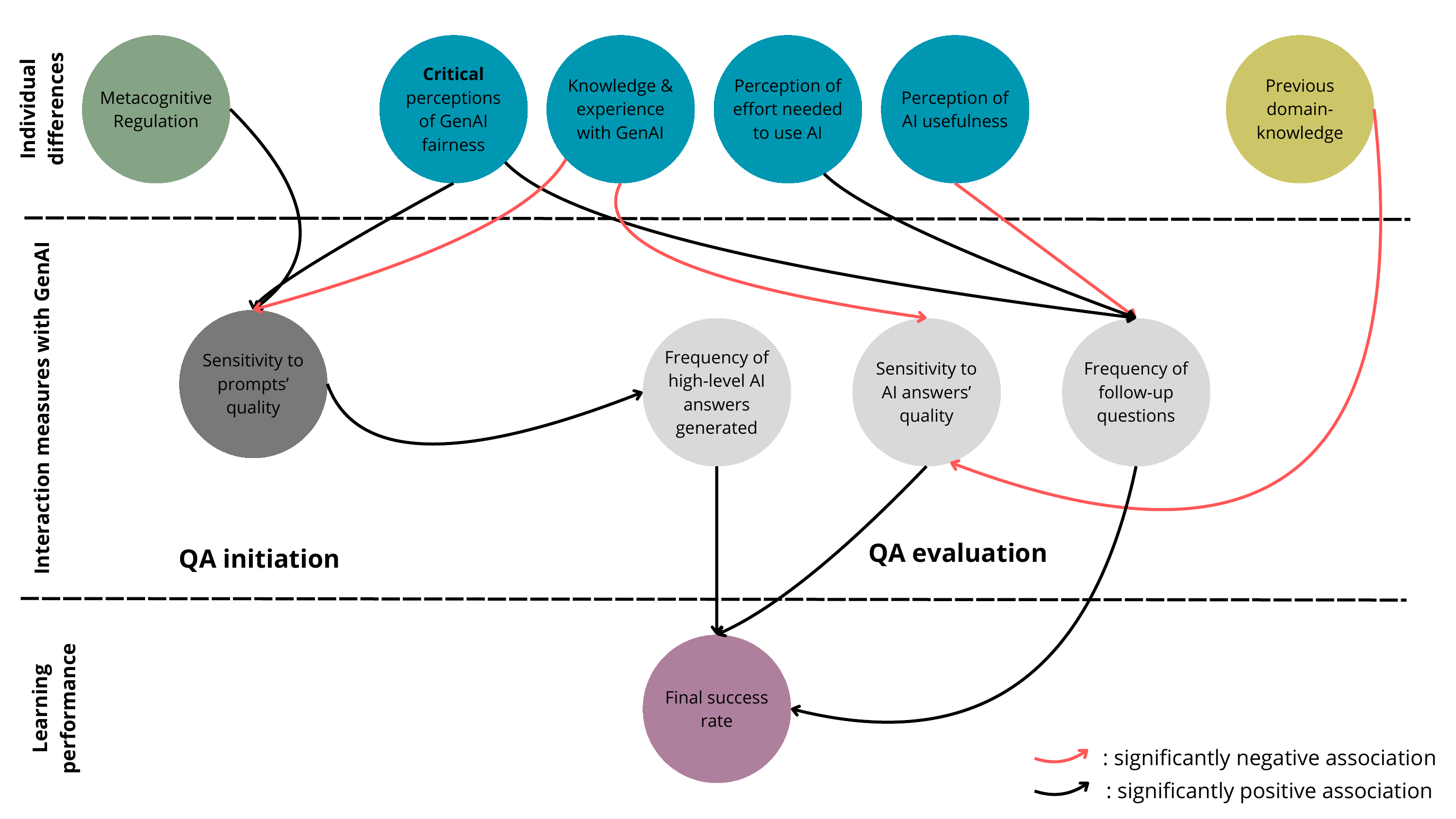}  \caption{\centering Summary of links between learning, LLM interaction behaviors, and individual differences}
\label{fig:summ-result}
\end{figure}

\paragraph{Summary}
While high-level answers from AI are necessary to help students' learning with the tool, students ability to access and verify that information is undermined by their over-confidence with AI attitudes and prior knowledge, thus suggesting the presence of AI literacy misconceptions. Conversely, metacognition and "healthy skepticism" (including critical perceptions of fairness, usefulness and effort needed to use AI) appear to be protective factors that drive meaningful QA cycles with LLMs. See~\autoref{fig:summ-result} for a summary of these results. 

\section{Discussion}
Our findings indicate that middle-school students struggled to engage in meaningful QA cycles during spontaneous learning with LLMs. More importantly, we saw that students with higher self-reported AI familiarity demonstrated lower proficiency in selecting effective prompts and evaluating responses. Overall, performance approached chance level, highlighting a critical gap in AI Literacy—defined here not merely as technical familiarity, but also as the capacity to use this knowledge to seek specific epistemic needs and verify information efficiently~\cite{chiu2024define}. Together with these deficits, students also tended to end their QA cycles prematurely, asking the minimal number of questions per task, even after receiving low-level AI answers, ultimately leading them to low success rates in solving the problems.

These results align with adult-focused research~\cite{denny2024generative,zamfirescu2023johnny,cheng2025asking} but contrast with developmental literature indicating that children as young as seven can identify efficient questions that efficiently convey their epistemic needs~\cite{mills2011determining} and recognize mechanistic, low-level explanations by age ten~\cite{mills2019want}. Rather than viewing this as a deficit in learner capability (i.e., a failure to transfer skills to LLM environments), we argue this highlights a unique challenge posed by GenAI design. Unlike human tutors who provide cues all along the QA process~\cite{cheng2025asking}, ChatGPT’s authoritative, seamlessly-effortless, and sometimes misleading tone imposes a substantial metacognitive load without offering counterpart scaffolding or opportunities for reflection on current information-search process~\cite{tankelevitch2024metacognitive,kidd2023}. This absence of support is likely exacerbated by the lack of pedagogical and metacognitive alignment in current LLMs~\cite{sonkar2024pedagogical,wang2025decoupling}. Such settings can lead to hampering the value of the QA process for learning for two reasons: 1) it discourages the cognitive investment required to evaluate answers, fostering over-reliance. Indeed, novices often perceive these systems as effortless and trustworthy, leading to reduced engagement during the interaction. 2) It imposes additional cognitive and metacognitive load on students, leaving little to no resources left for the germane processing of the task~\cite{stadler2024cognitive}, thus compromising the usefulness of the QA process for learning~\cite{Seufert2018TheLoad}. Our results further show that prior domain knowledge failed to mitigate these risks, leaving students vulnerable to inefficient inquiry even in subjects they understand~\cite{fazio2015knowledge,zhai2024effects}. 

Having established current usage patterns, we analyzed the role of individual differences, offering new insights for AI literacy frameworks. For instance, we identify metacognition as the missing link between AI "usage" and "competency"~\cite{kong2024developing}. Current instruction often focuses on technical mechanics or ethics of AI. However, our findings show these are insufficient without regulation: students with higher regulation scores were better able to judge prompt efficiency. This suggests that efficient AI Literacy represents a fundamentally Self-Regulated Learning (SRL) challenge: beyond retrieving accurate information, students must learn to leverage both domain and AI knowledge to actively regulate the goal of their QA-based LLM interactions toward their informational needs.

Consequently, training students to interact with LLMs seems to require shifting focus from purely technical aspects to metacognitive scaffolding. We propose two key shifts: 

\begin{itemize}
    \item \textbf{1) Training "Pedagogical Friction" and Effort Heuristics:} Interventions should expose students to detectable AI failures to recalibrate trust and encourage critical "friction" in the workflow. Students should be also introduced to differences between high- and low-level information (i.e. explanations that clarify concepts and helps understanding vs. isolated information) and why they should focus more on leveraging LLMs to seek the former. Here, we suggest that AI literacy frameworks must evolve to account for learning situations where the risk is not just factual error, but the illusion of explanatory depth—a critical distinction for science education which relies on evidence evaluation~\cite{chin2008students}.

Crucially, our finding that the \textit{perception of effort} correlated with the frequency of follow-up questions offers a tangible pedagogical lever. Since perceived effort was associated with increased QA regulatory behavior (through strategic follow-up), we suggest that this subjective feeling can serve as a valid metacognitive signal of active self-regulation. For instance, educators could explicitly teach students to interpret this friction positively: feeling cognitive effort while "fine-tuning" an exchange is a heuristic for effective QA. Conversely, students should be trained to view an "effortless" or frictionless interaction with suspicion, treating it as a signal to trigger epistemic vigilance. Furthermore, frameworks must evolve to account for the risk of the "illusion of explanatory depth"—teaching students to distinguish between high-level conceptual clarifications and isolated facts~\cite{chin2008students}.

\item \textbf{2) LLM Pedagogical Alignment by Design:} Beyond classroom interventions, these findings imply a need for LLM design changes. Future systems should move beyond passive answering or just giving cognitive cues to incorporate active metacognitive scaffolding (e.g., prompting users to clarify goals, formulate more precise questions, encourage checking one's understanding after each answer, etc) that guides meaningful QA cycles~\cite{sonkar2024pedagogical}.    
\end{itemize}
 
Taken together, the results in this work provide insights into the reality of student's low-quality QA cycles with LLMs. It encourages the need to implement targeted solutions, as these sub-optimal behaviors can lead to hindering their QA skills and epistemic vigilance on the long-term, and even standard learning settings~\cite{abd2023large,koos2023navigating}. Such interventions should not only focus on "How AI works", but also train their AI-based SRL strategies required to control it.

\section{Limitations and future directions}
A notable limitation of this study is the small sample size. Future studies addressing these limitations would benefit from larger sample sizes to enhance the robustness and generalizability of the findings. Additionally, due to time constraints, we were unable to collect data on important factors such as students' typing proficiency, linguistic skills, and fluency in asking questions in non-GenAI environments. These variables could provide a more comprehensive understanding of the factors influencing students' interaction with GenAI tools. 

Crucially, our study would also benefit from a more in-depth analysis of the students' self-generated questions, focusing on aspects such as linguistic quality, inclusion of necessary contextual elements, and the clarity of the instructions provided to ChatGPT. In this work, our evaluation of prompting skills relies mostly on students' selection of relevant questions from a pool of suggestions and the quality of the resulting answers. These indicators do not offer a comprehensive assessment of their prompting abilities. A more detailed analysis of self-generated prompts could thus give a valuable insights into students' prompting behaviors. By examining how they formulate their questions, we could better understand the specific challenges they face and identify areas for targeted improvement as well as the skills needed for their development. Using LLM-based deductive coding strategies could help us in this future analysis~\cite{xiao2023supporting}.

\section{Conclusion}
This study highlights a critical disconnect in middle school students' GenAI interaction:there is an over-reliance leading to inefficient inquiry and premature termination of QA learning cycles. These findings underscore the crucial need for interventions that go beyond technical mechanics to foster meaningful interactions. We urge educators to proactively integrate GenAI using strategies that induce "pedagogical friction" and explicit metacognitive monitoring. Ultimately, transforming AI literacy from passive consumption into an active, self-regulated competency is essential for preparing students to navigate the opportunities and challenges of these tools effectively.

\section{Acknowledgments}
Authors thank all participant students and teachers in the study. Authors also thank Dr. Didier Roy, Lucas Spooner and Jeremy Perez for their crucial help in recruiting participants and collecting data. The first author also benefited from funding from the University of Bordeaux Mobility grant (UBGRS) and the EdTech startup EvidenceB.

\section{Conflict of interest}
{All authors certify that they have no affiliations with or involvement in any organization or entity with any financial interest or non-financial interest in the materials discussed in this manuscript.}

\section{Declaration of generative AI and AI-assisted technologies in the writing process}
The first author used LLM-based AI tools (Anthropic's Claude and Google's Gemini) during the proof-reading process to enhance the readability and conciseness of the manuscript. The author made sure to review the models' suggestions and edit them before writing the final version of the text. The authors take full responsibility for the current content of the manuscript.

\bibliographystyle{abbrv}
\bibliography{sample,ref}

\end{document}